\documentclass[preprint,12pt,authoryear]{elsarticle}

\usepackage[authoryear]{natbib}
\usepackage{amssymb}
\usepackage{amsmath}
\usepackage{graphicx}
\usepackage{subcaption}
\usepackage{booktabs} 
\usepackage{pifont} 

\journal{Petroleum Science} 

\begin{document}

\begin{frontmatter}

\title{Seismic resolution enhancement via deep Learning with Knowledge Distillation and Domain Adaptation} 

\author[a]{Hanpeng Cai}
\author[a]{Haonan Zhang}
\author[a]{Liyu Zhang}
\author[b]{Suo Cheng}

\affiliation[a]{organization={the School of Resources and Environment, University of Electronic Science and Technology of China},
            addressline={No. 2006, Xiyuan Avenue}, 
            city={Chengdu},
            postcode={611731}, 
            state={Sichuan},
            country={China}}

\affiliation[b]{organization={PetroChina Tarim Oilfield Company and Research and Development Center for Exploration and Development Technology of Ultra -Deep Complex Oil and Gas Reservoirs},
            addressline={Institute of Tarim Oilfield Company}, 
            city={Korla},
            postcode={841000}, 
            state={Xinjiang},
            country={China}}

\begin{abstract}

High-resolution processing of seismic signals is critical for enhancing the accuracy of subsurface geological characterization and improving the identification of thin-layer reservoirs. Conventional high-resolution algorithms can partially recover high-frequency information but often suffer from poor robustness, low computational efficiency, and neglect of inter-trace structural relationships. In addition, many deep learning-based approaches adopt end-to-end architectures that lack the incorporation of prior knowledge or disregard disparities across data domains, resulting in insufficient generalization capability. Therefore, this paper proposes a deep learning network for seismic data high-resolution processing that integrates a knowledge distillation strategy with a domain adaptation mechanism, which is called the Domain-Adaptive Knowledge Distillation Network (DAKD-Net). Trained on datasets constructed through forward modeling, the network establishes physically constrained relationships between low and high resolution data. It extracts high-frequency prior knowledge during a guided phase before performing detail restoration without prior conditions. Domain adaptation further generalizes the model to practical seismic data, significantly enhancing its generalization capability and structural expression accuracy in real-field data. Structurally, DAKD-Net employs a U-Net backbone to fully extract spatial structural information from multi-trace seismic profiles. In its training mechanism, knowledge distillation enables prior knowledge transfer, allowing the network to recover high-resolution data directly from low-resolution inputs without requiring prior knowledge. For application, domain-adaptive fine-tuning substantially improves the network's generalization capacity and structural expression in actual survey areas. Experimental results demonstrate that the proposed method outperforms traditional approaches and classical deep networks in longitudinal resolution and the restoration of complex structural details, exhibiting strong robustness and practicality.

\end{abstract}




\end{frontmatter}

\section{Introduction}

\label{sec:introduction}

In seismic exploration, the quality of seismic data directly impacts the analysis and interpretation of subsurface geological structures. High-resolution seismic data offer detailed representations of subsurface features, enabling more accurate reservoir characterization and stratigraphic analysis. However, acquiring high-resolution seismic data in practical field operations is often challenging due to various limitations, and in most cases, only low-resolution data can be collected. Such low-resolution data are typically insufficient to reveal structural and lithological information effectively, leading to reduced exploration accuracy. Therefore, developing techniques to process low-resolution seismic data with high precision and efficiency, and to reconstruct high-resolution representations, remains a critical issue in seismic data processing.

Traditional high-resolution seismic data processing techniques include spectral whitening, deconvolution, and inverse Q filtering. Spectral whitening aims to compensate for frequency attenuation by broadening the amplitude spectrum of the signal. Common approaches include wavelet-based spectral whitening \citep{Chen2000improving}and Hilbert spectral whitening \citep{yang2007seismic}. Although these methods enhance high-frequency components, they suffer from limitations in phase and frequency-domain processing. Deconvolution, which mitigates the filtering effect caused by seismic wave propagation, is one of the primary techniques for improving seismic resolution. Representative methods include predictive deconvolution \citep{robinson1967predictive}, homomorphic deconvolution \citep{oppenheim1965superposition}, and spectral simulation deconvolution \citep{Zhao1996spectral}. To address the shortcomings of traditional spectral whitening and deconvolution approaches, various advanced methods have been proposed, such as sparse deconvolution based on L1 and Lp norms \citep{debeye1990lp}, high-resolution techniques based on the Hilbert-Huang transform \citep{dragomiretskiy2013variational}, sparse reflectivity inversion using compressed sensing \citep{Chen2015research}, generalized S-transform-based methods \citep{tian2009enhancing}, and wavelet estimation via neural networks for deconvolution \citep{lu1996neural}. The inverse Q-filtering algorithm was first proposed by Hale \citep{hale1981inverse}. Although stable and effective inverse Q-filtering algorithms and time-frequency domain methods \citep{wang2006inverse} circumvent the need for direct Q-value estimation, inverse Q-filtering still suffers from high computational complexity and poor noise immunity. These seismic high-resolution methods can partially restore high-frequency information but typically process single-trace data in the frequency or time-frequency domain. This approach neglects the continuity of inter-trace geological structures and holistic textural features. Moreover, their limited noise robustness and adaptability to complex geological settings fail to meet current demands for large-scale, high-precision seismic interpretation.

Deep learning, a prominent subfield of machine learning, has achieved groundbreaking success in various domains such as image recognition and natural language processing. Classical deep learning architectures include Generative Adversarial Networks (GANs) \citep{goodfellow2014generative}, Convolutional Neural Networks (CNNs) \citep{yuan2021improving}, and Residual Networks \citep{zhou2021seismic}. In recent years, deep learning has also shown promising advancements in geophysical applications, including lithology classification \citep{yu2019deep}, fault detection \citep{wu2019faultseg3d}, and fracture prediction \citep{gao2021channelseg3d}. The success of deep learning in image super-resolution has inspired its adoption in seismic high-resolution processing. By transforming low-resolution seismic data into high-resolution signals, these methods aim to enhance spatial resolution and improve the delineation of subsurface structures. Among these, CNNs are widely employed due to their ability to extract and reconstruct spatial features. The Super-Resolution Convolutional Neural Network (SRCNN), the first CNN-based model for image super-resolution, consists of three convolutional layers responsible for feature extraction, nonlinear mapping, and high-resolution reconstruction \citep{dong2014learning}. In 2019, Yuan et al. \citep{yuan2019enhanced} introduced CNNs into seismic data processing and proposed a CNN-based reconstruction model for seismic imaging. Building on this, Li et al. \citep{li2021deep} developed a deep convolutional network that simultaneously performs seismic image super-resolution and denoising. 

In addition, numerous other models have been introduced for high-resolution tasks. Ledig et al. \citep{ledig2017photo} were the first to apply GANs to high-resolution processing, proposing a high-resolution generative adversarial network model. Lan et al. \citep{lan2023simultaneous} achieved noise suppression and resolution enhancement of seismic data using a method based on elastic convolution dictionary learning. In 2024, Zhang et al. \citep{zhang2024seisresodiff} proposed a diffusion model-based seismic data resolution enhancement method. U-Net has also been widely applied in image high resolution. Sun et al. \citep{sun2021end} utilized a U-Net model to train pseudo-reflectivity models for end-to-end seismic resolution enhancement. Li et al. \citep{li2023time} further proposed a method that integrates the S-transform and complex-valued U-Net to jointly learn both time-domain and frequency-domain features from seismic data. However, most existing deep learning methods operate within end-to-end frameworks that directly learn mapping relationships between low-resolution and high-resolution seismic data. They lack constraints from seismic prior knowledge, and frequently rely on synthetic training data. These limitations result in insufficient generalization capability for real seismic data, hindering deployment in diverse field environments.

Knowledge distillation is a teacher-student network paradigm that enables the transfer of knowledge from a large, high-capacity teacher model to a smaller, more efficient student model. This approach aims to reduce computational complexity and memory usage while maintaining comparable model performance. First introduced by Hinton et al. in 2015 \citep{hinton2015distilling}, knowledge distillation has since been widely adopted in various scenarios such as model compression, transfer learning, and multi-task learning. The core concept of knowledge distillation lies in having the student model learn not only from the ground-truth labels (hard labels). It also learns from the teacher model's output probabilities (soft labels), which encode additional information about the data distribution and inter-class relationships. To leverage prior knowledge in high-resolution reconstruction tasks, Wang et al. \citep{wang2022propagating} proposed a knowledge distillation framework for face recognition. Their approach enables the student network to implicitly learn prior knowledge without explicit estimation. Specifically, the teacher model is trained with access to accurate prior information, and the distilled knowledge is transferred to the student model. As a result, the student model can directly reconstruct high-resolution face images from low-resolution inputs, effectively mitigating the adverse effects caused by inaccurate prior estimation.

To address the issue of insufficient utilization of structural information between seismic traces in traditional methods such as spectral whitening, as well as the inadequate generalization capabilities of some end-to-end networks, we propose a deep learning-based seismic high-resolution processing network that integrates knowledge distillation and domain adaptation—DAKD-Net. First, a training dataset is constructed through the simulation of the seismic forward modeling process, which includes low-resolution input data, high-resolution prior data, and high-resolution label data. During the training phase, high-resolution priors are introduced to achieve precise reconstruction of high-frequency details. Concurrently, soft label loss and feature space loss are employed to guide the network in learning and transferring prior knowledge. In the practical data phase, a domain adaptation strategy is utilized to fine-tune the network's high-level feature modules, enabling efficient migration of the model from the synthetic domain to the real seismic domain, thereby enhancing the network's generalization ability for real seismic data. Experimental validation demonstrates that DAKD-Net not only excels in model data but also exhibits remarkable improvements in resolution and structural identification capabilities in actual worksite scenarios.

\section{Methods}
\label{sec:methods}

The objective of the DAKD-Net proposed in this paper is to automatically recover high-resolution features from low-resolution seismic data without the need for direct input of prior information. Its overall architecture is shown in Fig. 1 and consists of three core modules: feature extraction module, guided learning module, and domain-adaptive transfer module. The network is built on a U-Net architecture, utilizing a symmetric encoder-decoder structure and skip connections to extract and reconstruct cross-layer information, fully considering the structural continuity and texture features between channels. The overall method is divided into three stages:
Guided training stage: By constructing a synthetic seismic dataset, a physical mapping relationship between low- and high-resolution data is established, and prior data is introduced to guide the network in learning the structural representation of high-frequency features. In this stage, the feature extraction module extracts intermediate features at multiple scales, and provides supervision signals for subsequent stages through soft labels and feature space loss.

Autonomous recovery stage: The prior input is removed, and the guided learning module relies solely on low-resolution data to recover high-frequency features. During training, the network's ability to restore detailed features is enhanced through intermediate feature supervision and structural similarity loss from the guided training stage, enabling it to achieve high-resolution reconstruction without prior conditions.

Domain Adaptive Transfer Stage: Based on a small amount of actual seismic data, the lower-level feature extraction module is frozen, and only the parameters of the higher-level components are fine-tuned to achieve transfer from the synthetic domain to the actual seismic data domain, thereby improving the network's generalization capability in real-world data.

\begin{figure}[htbp]
  \centering
  \includegraphics[width=\textwidth]{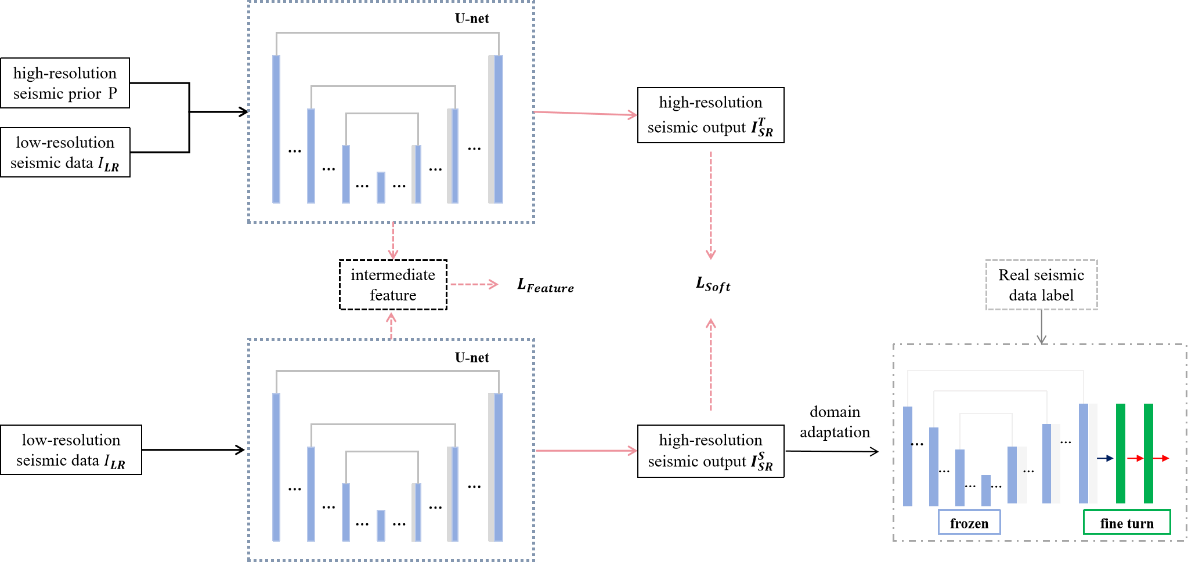} 
  \caption{Architecture of the Knowledge Distillation-based Seismic High-Resolution Network.} 
  \label{fig:architecture of knowledeg} 
\end{figure}

\subsection{Unet}
\label{sec1}
U-Net is a fully convolutional neural network architecture specifically designed for pixel-level segmentation tasks. It features a symmetric encoder-decoder structure with skip connections, enabling efficient and accurate image segmentation across various applications. The U-Net consists of two main components: an encoder and a decoder \citep{ronneberger2015u}. The network adopts a U-shaped architecture. The encoder progressively downsamples the input to extract deep features, while the decoder upsamples the feature maps to reconstruct the output. A key feature of U-Net is the incorporation of "skip connections" that directly link feature maps of the same resolution between the encoder and decoder. These skip connections transmit shallow features to the decoding path, helping to preserve high-resolution information that might otherwise be lost during downsampling. With its symmetric design and skip connections, U-Net effectively facilitates multi-scale feature fusion.

The architecture of the U-Net network is illustrated in Fig. 2. The number of feature channels starts at 64 and doubles progressively up to 512. In the encoder part of the U-Net, each layer consists of two convolutional layers with a kernel size of 3*3 and a stride of 1 to preserve the spatial dimensions. Each convolutional layer is followed by a ReLU activation function to introduce non-linear feature representation capabilities. A 2*2 max-pooling layer is then applied for downsampling, which reduces the spatial resolution. At the same time, it doubles the number of channels, enabling the extraction of increasingly abstract features while gradually losing fine details. In the decoder part, transposed convolutions are used for upsampling to progressively restore the spatial resolution. The upsampled feature maps are concatenated with the corresponding encoder feature maps via skip connections along the channel dimension. After concatenation, the feature maps pass through two additional 3*3 convolutional layers, each followed by a ReLU activation function. As the resolution is restored layer by layer, the number of feature channels is correspondingly reduced. Finally, a 1*1 convolutional layer is used to generate the output image.

\begin{figure}[htbp]
  \centering
  \includegraphics[width=\textwidth]{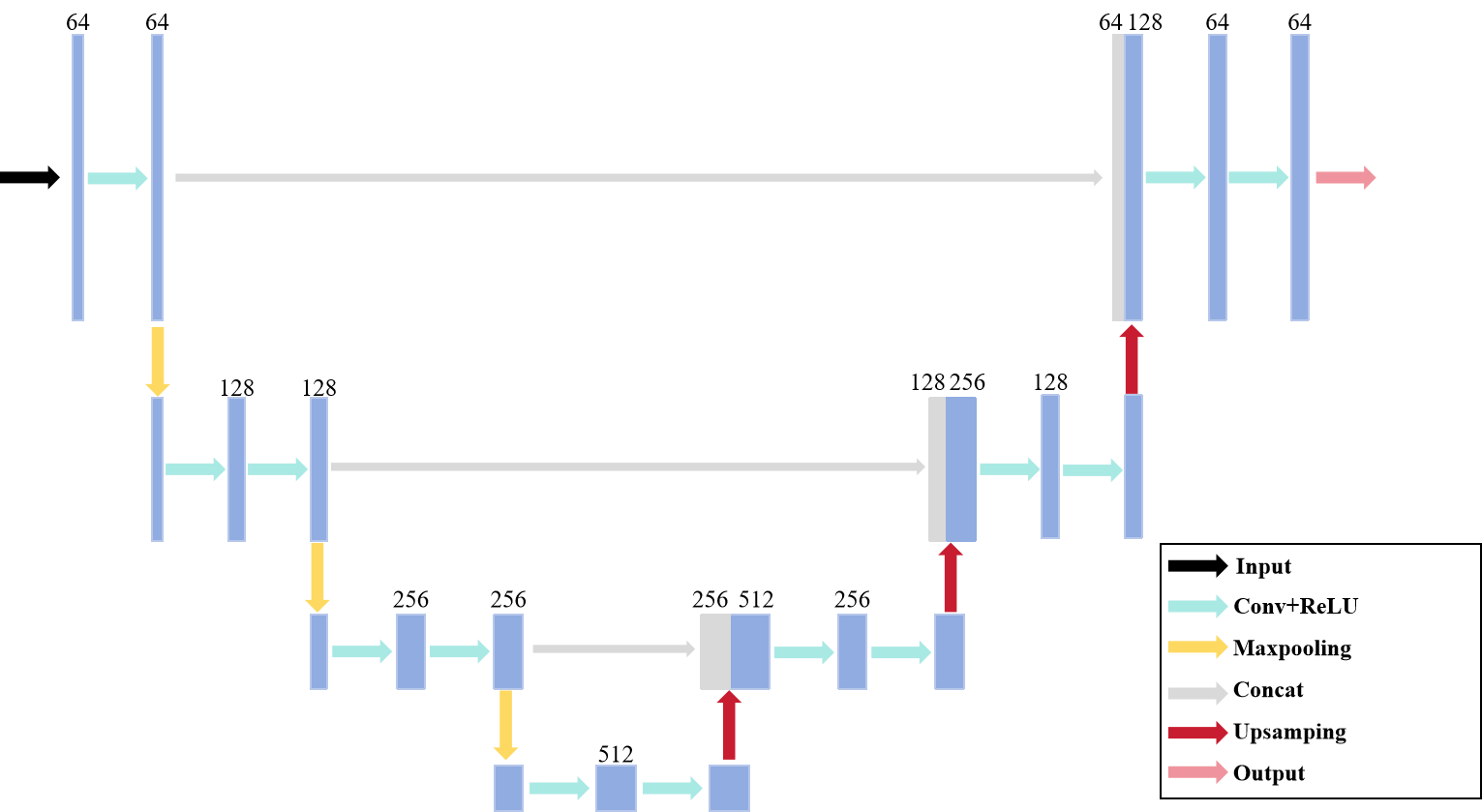} 
  \caption{Architecture of  U-net network.} 
  \label{fig:architecture of Unet} 
\end{figure}

\subsection{Feature Extraction Module}
\label{sec2}

Conventional CNN-based high-resolution methods often suffer from blurred reconstructions and loss of detail in the absence of geological structural information. This issue becomes more pronounced in regions with complex geological features such as faults and structural boundaries. Therefore, incorporating seismic data prior information into the feature extraction module enables the network to learn more accurate geological structure and detail information, which is crucial for improving high-resolution performance. By using low-resolution seismic data and high-resolution prior knowledge as joint inputs, and concatenating them before inputting into the network, the process is defined as:

\begin{equation}
I_{SR}^T = f_{\text{Feature}} \left( \left[ I_{LR}, P \right] \right)
\end{equation}

where $P$ denotes the high-resolution seismic prior, $I_{LR}$ represents the low-resolution seismic data, $f_{\text{Feature}}$ represents the feature extraction module function, $I_{SR}^T$ is the high-resolution seismic output generated by this module, and $[ , ]$ denotes the concatenation operator.

The feature extraction module is only used during the guided training phase, at which point synthetic data is used for training, enabling the acquisition of accurate high-resolution prior information. When generating synthetic data, convolution is performed between subwaves of different frequency bands and reflection coefficients to produce low-, medium-, and high-frequency seismic data. Among these, the convolutional data of the mid-frequency wavelet is selected as the high-resolution prior for the following reasons: first, the key to high-resolution seismic data lies in its spectral information, and the mid-frequency wavelet can directly provide the detailed frequency characteristics required for spectral extension, containing rich structural details in the seismic data; second, using the prior information synthesized from the mid-frequency wavelet can provide more universal frequency characteristics, avoiding the model's over-reliance on specific geological interpretations. 

The specific architecture of the feature extraction module based on the U-net network is shown in Fig. 2, which includes the following key parts:

\begin{enumerate}
  \item \textnormal{Input Layer:} The low-resolution seismic data $I_{LR}$ and high-resolution prior information $P$ are concatenated as joint input to the network.
  \item \textnormal{Feature Extraction Module:} Convolutional layers are used to extract deep features. Features $F_P$ are extracted from the high-resolution prior via a convolutional path, while features $F_{LR}$ are extracted from the low-resolution seismic data using a combination of convolutional and upsampling layers. Due to the limited spatial resolution of the low-resolution input, direct use in downstream processing may result in insufficient structural representation. Thus, upsampling is applied to compensate for spatial resolution loss and ensure compatibility during feature fusion.
  \item \textnormal{Feature Fusion Module:} A concatenation layer is used for feature fusion. The depth features extracted from high-resolution prior $F_P$ and low-resolution data $F_{LR}$ are concatenated to generate a joint feature tensor $F_{\text{joint}}$. This joint representation captures both the reflection patterns from the low-resolution input and the structural details provided by the high-resolution prior.
  \item \textnormal{Encoder-Decoder Structure:} The fused feature tensor $F_{\text{joint}}$ is fed into the U-Net architecture. The encoder progressively downsamples the features to capture multi-scale contextual information, while the decoder restores spatial resolution using transposed convolutions. Skip connections bridge encoder and decoder stages to preserve spatial details.
  \item \textnormal{Residual Learning Module:} To further enhance high-frequency detail reconstruction, a residual connection is added before the final output. The predicted high-resolution result is added to the original low-resolution input, enabling the network to focus on learning the missing high-frequency components.
  \item \textnormal{Output Layer:} The network outputs the predicted high-resolution seismic data $I_{SR}^T$. Intermediate feature maps from each layer are retained for use during the knowledge distillation phase.
\end{enumerate}

\begin{figure}[htbp]
  \centering
  \includegraphics[width=\textwidth]{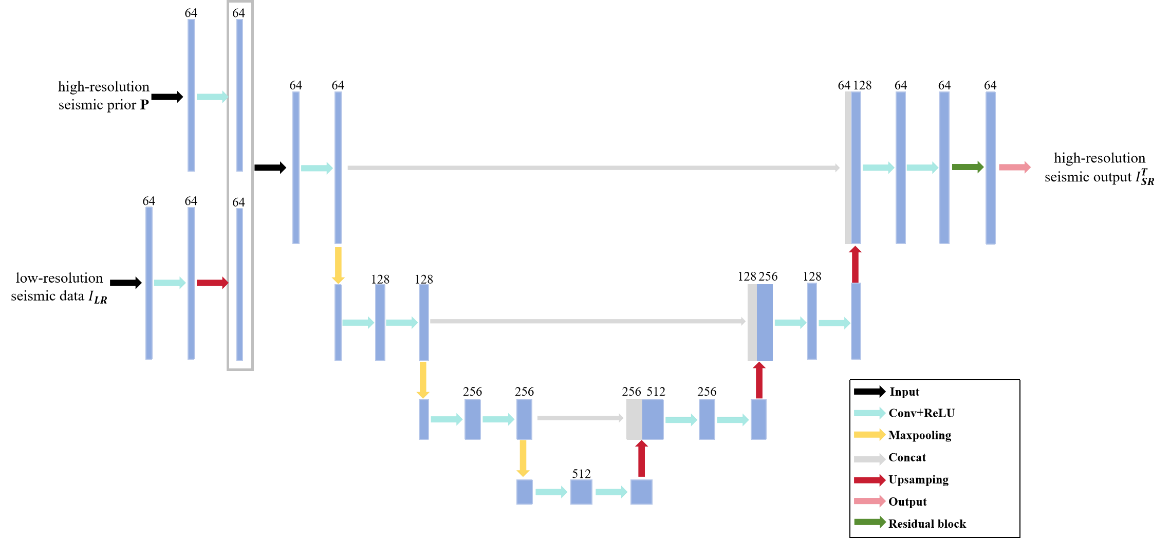} 
  \caption{Feature extraction module of DAKD-Net.} 
  \label{fig:feature of DAKD} 
\end{figure}

To constrain the feature extraction module to generate high-resolution seismic data, a loss function combining $L_1$ loss and Structural Similarity (SSIM) loss is adopted. The SSIM loss emphasizes the preservation of textural details and structural similarity within seismic data. By incorporating the SSIM component, the model is encouraged to maintain fine-grained features such as seismic reflections and geological boundaries. This helps enhance the fidelity of the reconstructed seismic signals. The $L_{\text{Voxel}}$ loss is defined as:

\begin{equation}
L_{\text{Voxel}} = \left\| I_{SR}^T - I_{HR} \right\|_1
\end{equation}

where $I_{HR}$ denotes the input high-resolution seismic data.

SSIM is defined as:

\begin{equation}
SSIM(x, y) = \frac{(2\mu_x\mu_y + C_1)(2\sigma_{xy} + C_2)}{(\mu_x^2 + \mu_y^2 + C_1)(\sigma_x^2 + \sigma_y^2 + C_2)}
\end{equation}

Here, $x$ and $y$ represent two seismic data samples, $\mu_x$ and $\mu_y$ denote their respective means, $\sigma_x^2$ and $\sigma_y^2$ are the standard deviations, and $\sigma_{xy}$ is the covariance between $x$ and $y$. Constants $C_1$ and $C_2$ are stabilization terms used to avoid division by zero. A higher SSIM value indicates a greater structural similarity between the predicted high-resolution result and the high-resolution ground truth. Therefore, $L_{\text{SSIM}}$ is defined as $1 - SSIM(x, y)$, aiming to minimize structural discrepancies between seismic data, which is defined as:

\begin{equation}
L_{\text{SSIM}} = 1 - SSIM(I_{SR}^T, I_{HR})
\end{equation}

The final loss function for the feature extraction module is defined as:

\begin{equation}
L_{\text{Teacher}} = \alpha \times L_{\text{Voxel}} + (1 - \alpha) \times L_{\text{SSIM}}
\end{equation}

where $\alpha$ is the weight of $L_{\text{Voxel}}$ and $L_{\text{SSIM}}$. During the training process, the weights of the $L_1$ loss and SSIM loss are dynamically adjusted. In the early stages, greater emphasis is placed on the loss to ensure numerical accuracy. In the later stages, the weight of the SSIM loss is gradually increased to enhance structural similarity. This strategy ensures a balanced trade-off between numerical precision and structural consistency in the model's output.

\subsection{Guided Learning Module}
\label{sec3}
To enhance the network's high-frequency recovery capability under conditions without prior information input, DAKD-Net adopts a “guided-then-autonomous” training mechanism. During the guided training phase, the network jointly uses low resolution seismic data and precise prior data at the input end, and trains using a dataset synthesized through forward modeling, effectively obtaining geological structural information. However, in actual seismic exploration, it is challenging to obtain complete and accurate prior information about faults. Therefore, in the autonomous recovery phase, prior information is removed, and only low-resolution seismic traces are used as input. During training, the network continuously receives soft label supervision and feature space distillation information from the guided training phase, achieving generalization and transfer from “with prior information” to “without prior information.” This mechanism enables the network to independently reconstruct high-resolution seismic data during the inference stage while maintaining structural consistency. The two-stage collaborative training strategy enhances the network's ability to express complex geological features through constraints in the structural space.

The structure of the guided learning module is shown in Fig. 4. It is similar to the structure of the feature extraction module, but there is a key difference in terms of input: the guided learning module only receives low-resolution seismic data as input and does not require additional prior information. In addition, the guided learning module outputs features at each decoding stage and compares them with the outputs of the corresponding stages of the feature extraction module to ensure that the detailed knowledge in the feature extraction module can be learned at different scales.

\begin{figure}[htbp]
  \centering
  \includegraphics[width=\textwidth]{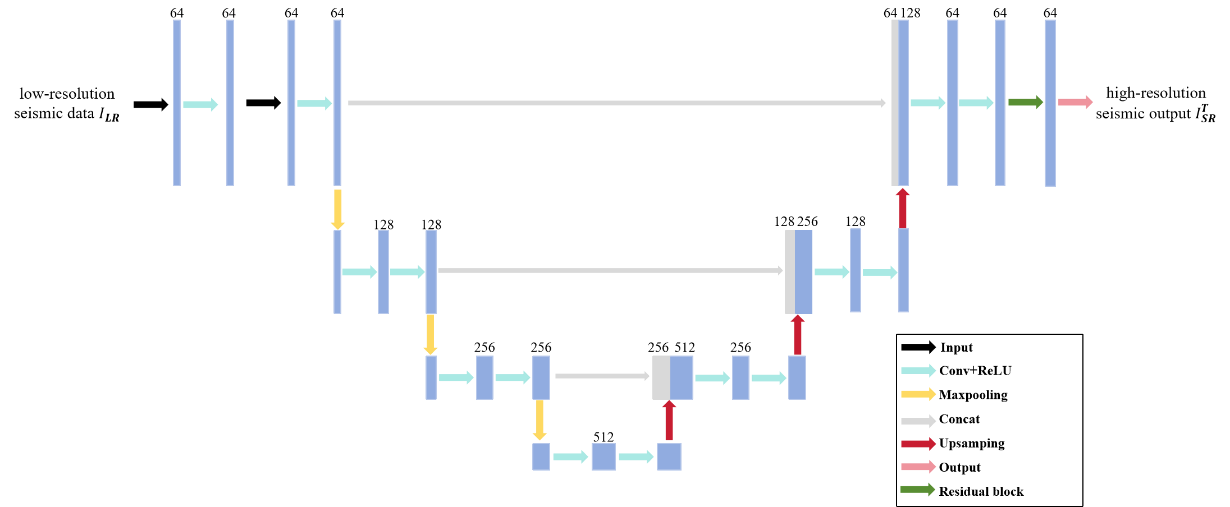} 
  \caption{Guided learning module of DAKD-Net.} 
  \label{fig:guide of DAKD} 
\end{figure}

Through the knowledge distillation strategy, the intermediate features output by the feature extraction module are passed to the guided learning module, which compares its own intermediate layer features with them to minimize the differences between the two. The loss function of the guided learning module consists of multiple loss terms, including  loss, SSIM loss, soft label loss, and feature space loss, defined as follows:

\begin{equation}
  L_{\text{Guided}} = \alpha \times L_{\text{Pixel}} + \beta \times L_{\text{Feature}} + \gamma \times L_{\text{Soft}} + \kappa \times L_{\text{SSIM}}
\end{equation}

where $\alpha$, $\beta$, $\gamma$, and $\kappa$ are the weights of $L_{\text{Pixel}}$, $L_{\text{Feature}}$, $L_{\text{Soft}}$, and $L_{\text{SSIM}}$, respectively. To effectively transfer the high-resolution seismic prior from the feature extraction module to the guided learning module, two knowledge distillation losses are introduced: the soft ground truth loss and the feature space loss. Since relying solely on pixel-wise differences may be insufficient to capture critical structural features in seismic data, the SSIM loss is also incorporated to enhance the structural fidelity of the generated data. SSIM loss evaluates the local structural similarity between seismic signals. It ensures that the high-resolution outputs generated by the student network are not only numerically close to the reference data but also consistent in terms of seismic waveform characteristics and geological structures.

$L_{\text{Pixel}}$ is the pixel-wise $L_1$ loss function applied on the $I_{SR}^S$ and $I_{HR}$, which is defined as:

\begin{equation}
L_{\text{Pixel}} = \left\| I_{SR}^S - I_{HR} \right\|_1
\end{equation}

where $I_{SR}^S$ denotes the high-resolution prediction produced by the guided learning module, and $I_{HR}$ represents the high-resolution seismic ground truth.

The high-resolution prediction results generated by the feature extraction module are referred to as soft labels, while the actual high-resolution data are referred to as hard labels. Soft labels can include varying degrees of fuzziness and uncertainty, making the learning objectives more broad and thereby improving adaptability and generalization capabilities for unknown seismic data, providing simpler but more effective supervision for the network. In order to make the output of the autonomous recovery stage closer to the results of the guided training stage, soft ground truth loss $L_{\text{Soft}}$ is used, which is implemented based on the $L_1$ loss, as shown in the following formula:

\begin{equation}
L_{\text{Soft}} = \left\| I_{SR}^S - I_{SR}^T \right\|_1
\end{equation}

where $I_{SR}^S$ represents the high-resolution prediction results of the autonomous recovery stage, and $I_{SR}^T$ represents the prediction results of the guided training stage, namely the soft ground truth, which is generated from $I_{LR}$ and $P$. The soft ground truth contains rich high-resolution prior information and is easier to learn compared to the ground truth targets.

To explicitly constrain the intermediate features between the guided training phase and the autonomous recovery phase, the $L_{\text{Feature}}$ feature space loss is used to calculate the difference between the output features of the two in the intermediate layer, as shown in the following formula:

\begin{equation}
L_{\text{Feature}} = \frac{1}{n} \sum_{i}^{n} \left\| F_i^S - F_i^T \right\|_1
\end{equation}

where $F_i^S$ and $F_i^T$ represent the intermediate layer features of the guided training stage and the autonomous recovery stage, respectively. These features are typically the output tensors of intermediate layers, such as activations following convolutional layers. $n$ represents the number of selected intermediate layers for comparison. Through the feature space loss, the guided learning module can learn the deep seismic structural information contained in the feature extraction module without prior knowledge, thereby improving the detail representation capability of the reconstruction.

\subsection{Domain Adaptation}
\label{sec4}

Due to significant differences between synthetic seismic data and actual seismic data in terms of wave morphology, frequency characteristics, and geological structure, directly applying a trained model to an actual work area may encounter issues. For instance, differences between actual seismic data and synthetic data in subwaveform morphology and geological structure may render the model unable to adapt to new data; or, actual work area data is often limited, making it difficult to support training a complex deep learning model from scratch. Even if the network can be trained from scratch, it would still require a significant amount of time and computational resources \citep{pan2010survey}. To address this, DAKD-Net introduces a domain adaptation mechanism to enable effective model transfer from the synthetic domain to the real domain.

Domain adaptation can, on one hand, fine-tune the model to adapt to the differences between synthetic data and actual work area data, and on the other hand, significantly reduce training time by transferring existing model weights \citep{Tang2021fault}. Additionally, in cases where actual work area data is insufficient, transfer learning can significantly reduce reliance on large amounts of labeled data. Therefore, after initially training the high-resolution seismic data network DAKD-Net using synthetic data, the pre-trained network's weights are transferred to the new task network, with the lower-level features of the network frozen and only the parameters of the higher-level parts of the network fine-tuned. A small amount of actual work area data is used to update part of the pre-trained model's parameters, enabling the model to better adapt to the characteristics of actual work area data. This strategy effectively mitigates the “domain shift” issue, enabling the model to possess the strong constraint learning capabilities from synthetic data while adapting to the characteristics of field seismic data, thereby enhancing its generalization ability and application scope.

\subsection{High-Resolution Evaluation Metrics}
\label{sec5}

For the output results of the high-resolution network, two quantitative metrics are used for evaluation: Peak Signal-to-Noise Ratio (PSNR) and SSIM, which comprehensively assess the quality of the output from the perspectives of data fidelity and structural similarity, respectively.

PSNR is a commonly used metric for measuring image reconstruction quality. It is calculated based on the ratio of the maximum possible signal power to the noise power of the signal. PSNR is often used to assess the quality of an image after high-resolution processing. A higher PSNR value indicates that the reconstructed image has less difference from the original image, resulting in higher image quality. The formula for calculating PSNR is as follows:

\begin{equation}
PSNR = 10 \log_{10} \frac{R^2}{\frac{1}{m \times n} \sum_{i=1}^{m} \sum_{j=1}^{n} \left( I_{SR}(i, j) - I_{HR}(i, j) \right)^2}
\end{equation}

where $R$ is the maximum possible pixel value of the image. For 8-bit images, $R=255$. $m \times n$ represents the size of the seismic image, $I_{SR}(i, j)$ represents the pixel values of the high-resolution predicted result, and $I_{HR}(i, j)$ represents the pixel values of the high-resolution label data.

SSIM is used to measure the structural similarity between the high-resolution predicted results and high-resolution label data, rather than just pixel-wise differences. Compared to PSNR, SSIM values are more aligned with human visual perception of image quality. The closer the SSIM value is to 1, the higher the structural similarity between the high-resolution predicted results and the high-resolution label data. Its definition is given by the earlier equation (3).

\section{Experiments and Discussion}
\label{sec:experiments}

\subsection{ Network Training and Validation}
\label{sec6}

The reflection coefficients of the Marmousi2 model are convolved with low-frequency and high-frequency wavelets to construct low-resolution seismic data as input data and high-resolution seismic data as labels. Additionally, middle-frequency band wavelets are randomly selected from low-frequency and high-frequency wavelets for convolution to serve as high-resolution prior information. To enhance the generalization ability of the network, low-frequency wavelets are randomly selected Ricker wavelets with a central frequency between 20-30Hz. High-frequency wavelets are selected Ricker wavelets with a central frequency of 80Hz, while middle-frequency band wavelets are selected Ricker wavelets with a central frequency of 40Hz, all with a sampling time of 1ms. Multiple sets of high-resolution data with a size of 1000*300, as well as corresponding low-resolution data and high-resolution priors, are constructed, as shown in Fig. 5. To improve the robustness of the network, Gaussian noise with different variances is randomly added to the low-resolution synthetic data. Subsequently, all synthetic data with a size of 1000*300 is divided into synthetic data with a size of 128*128 as training and validation datasets by setting an appropriate sliding step, as shown in Fig. 6. To improve the generalization ability of the network model and prevent overfitting, data augmentation operations such as random vertical and horizontal flips are applied after normalizing the synthetic data. The ratio of the training set to the validation set is set to 9:1.

\begin{figure}[htbp]
  \centering
  \begin{subfigure}[b]{0.32\textwidth}
    \includegraphics[width=\textwidth]{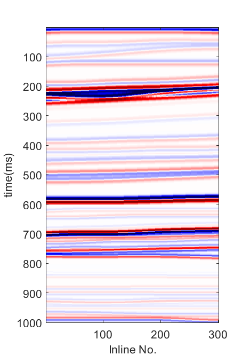}
    \caption{}
  \end{subfigure}
  \begin{subfigure}[b]{0.32\textwidth}
    \includegraphics[width=\textwidth]{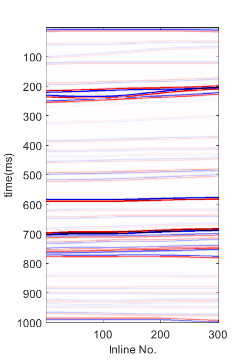}
    \caption{}
  \end{subfigure}
  \begin{subfigure}[b]{0.32\textwidth}
    \includegraphics[width=\textwidth]{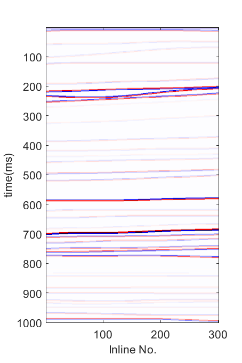}
    \caption{}
  \end{subfigure}
  \caption{Synthetic data of size 1000*30. (a) Low-resolution synthetic data; (b) High-resolution prior data; (c) High-resolution synthetic data.}
  \label{fig:synthetic_data}
\end{figure}

\begin{figure}[htbp]
  \centering
  \begin{subfigure}[b]{0.32\textwidth}
    \includegraphics[width=\textwidth]{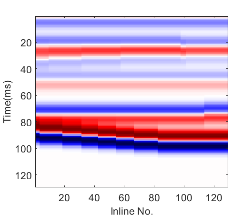}
    \caption{}
  \end{subfigure}
  \begin{subfigure}[b]{0.32\textwidth}
    \includegraphics[width=\textwidth]{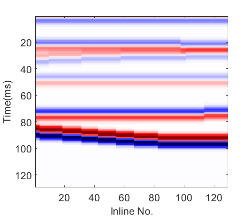}
    \caption{}
  \end{subfigure}
  \begin{subfigure}[b]{0.32\textwidth}
    \includegraphics[width=\textwidth]{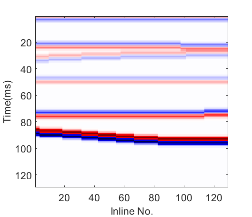}
    \caption{}
  \end{subfigure}
  \caption{Synthetic dataset. (a) Low-resolution synthetic data of size 128*128; (b) High-resolution prior data of size 128*128; (c) High-resolution synthetic data of size 128*128.}
  \label{fig:synthetic_dataset}
\end{figure}

Fig. 7 and Table I show the loss curves of the feature extraction module at different loss weights $\alpha$, along with the corresponding PSNR and SSIM values. These results indicate that when $\alpha=0.7$, the trade-off between SSIM loss and $L_1$ loss is most reasonable, with both PSNR and SSIM reaching high levels, and the best high-resolution processing results. The training parameters for the feature extraction module are set as follows: loss weight = 0.7, learning rate = 0.00001, optimizer = Adam optimizer, number of iterations = 30, and batch size = 16. During the self-recovery phase, the parameters of the feature extraction module remain frozen, with the intermediate layer features and high-resolution output serving as guidance information for the guidance learning module. The loss function $L_{\text{Guided}}$ for the student network is defined as follows, where $L_{\text{Pixel}}$ weight $\alpha = 1.4$, $L_{\text{Soft}}$ weight $\gamma = 0.1$, and $L_{\text{SSIM}}$ weight $\kappa = 0.4$. The $L_{\text{Feature}}$ weight needs to be dynamically adjusted. The $L_{\text{Feature}}$ weight $\beta$ is defined as:

\begin{equation}
  \beta =
  \begin{cases}
  1 - 0.2 \cdot \text{epoch}, & \text{if } \text{epoch} < d \\
  0, & \text{if } \text{epoch} \geq d
  \end{cases}
  \label{eq:beta}
\end{equation}

where epoch represents the number of iterations in the student network's training process, and $d$ is the threshold. When training the Marmousi2 model, $d = 5$. Similarly, the Adam optimizer is used with 30 iterations and a batch size of 16.

\begin{figure}[htbp]
  \centering
  \begin{subfigure}[b]{0.45\textwidth}
    \includegraphics[width=\textwidth]{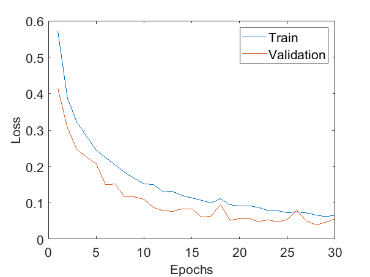}
    \caption{}
  \end{subfigure}
  \begin{subfigure}[b]{0.45\textwidth}
    \includegraphics[width=\textwidth]{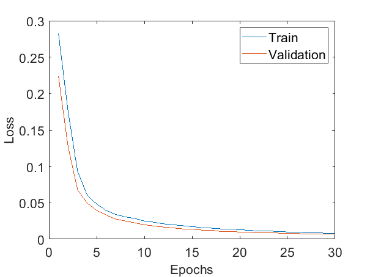}
    \caption{}
  \end{subfigure}
  \begin{subfigure}[b]{0.45\textwidth}
    \includegraphics[width=\textwidth]{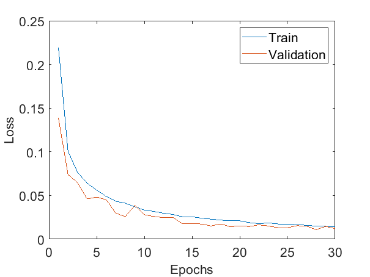}
    \caption{}
  \end{subfigure}
  \begin{subfigure}[b]{0.45\textwidth}
    \includegraphics[width=\textwidth]{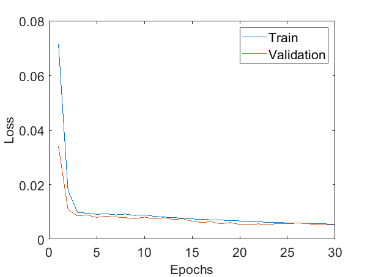}
    \caption{}
  \end{subfigure}
  \caption{Feature extraction module loss curves corresponding to different loss weights. (a) 0.5 loss weight; (b) 0.7 loss weight; (c) 0.9 loss weight; (d) 1.0 loss weight.}
  \label{fig:loss_curves}
\end{figure}

\begin{table*}[htbp] 
\renewcommand{\arraystretch}{1.5} 
\centering
\caption{SSIM and PSNR Corresponding to Different Loss Weights $\alpha$}
\label{tab:ssim_psnr}
\begin{tabular*}{\textwidth}{@{\extracolsep{\fill}}lcc} 
\toprule
$\alpha$ & SSIM   & PSNR   \\ 
\midrule
0.5      & 0.8964 & 32.8518 \\ 
0.7      & 0.9846 & 38.2939 \\ 
0.9      & 0.9398 & 33.2817 \\ 
1.0      & 0.9125 & 33.2485 \\ 
\bottomrule
\end{tabular*}
\end{table*}

\subsection{Synthetic Seismic Data}
\label{sec7}

To thoroughly evaluate the effectiveness of the DAKD-Net, synthetic datasets generated from the Marmousi2 model are used to train and test both the proposed high-resolution network and a traditional fully convolutional network (FCN).

Fig. 8 shows the loss curve of DAKD-Net during the training phase and the high-resolution evaluation metrics (PSNR and SSIM curves) on the validation set. It can be seen that the loss curve of DAKD-Net decreases steadily, and the PSNR and SSIM metrics both reach high levels. Additionally, the loss values of the training set and validation set are close, and the improvement process of PSNR and SSIM is consistent, indicating that DAKD-Net can effectively avoid overfitting and has good generalization ability.

\begin{figure}[htbp]
  \centering
  \begin{subfigure}[b]{0.32\textwidth}
    \includegraphics[width=\textwidth]{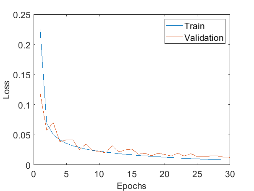}
    \caption{}
  \end{subfigure}
  \begin{subfigure}[b]{0.32\textwidth}
    \includegraphics[width=\textwidth]{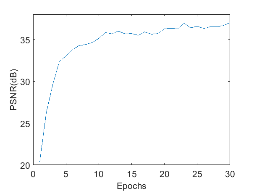}
    \caption{}
  \end{subfigure}
  \begin{subfigure}[b]{0.32\textwidth}
    \includegraphics[width=\textwidth]{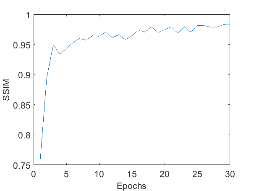}
    \caption{}
  \end{subfigure}
  \caption{Loss functions and evaluation metrics of DAKD-Net. (a) Loss function of DAKD-Net on the training and validation sets; (b) PSNR values of DAKD-Net on the validation sets; (c) SSIM values of DAKD-Net on the validation sets.}
  \label{fig:loss_metrics}
\end{figure}

Fig. 9 shows the high-resolution prediction results of DAKD-Net for the Marmousi2 model data. Fig. 9(a) is a low-resolution seismic section with a dominant frequency of 20 Hz, where the phase axis contours are blurred and details are unclear. Fig. 9(b) is a true high-resolution seismic section with a dominant frequency of 80 Hz, where structural details are more distinct. Fig. 9(c) shows the prediction results of DAKD-Net. It can be observed that DAKD-Net can refine the blurred details in the low-resolution profile while preserving the overall structure of the seismic profile, significantly enhancing the precision of the phase axis. In particular, complex geological structures become clearer in the prediction results, with a noticeable improvement in the precision of the phase axis compared to the low-resolution data, and an overall increase in resolution. The results also exhibit a high degree of similarity to the true high-resolution data in Fig. 9(b), indicating that DAKD-Net can effectively enhance the resolution of seismic data.

\begin{figure}[htbp]
  \centering
  \begin{subfigure}[b]{0.8\textwidth}
    \includegraphics[width=\textwidth]{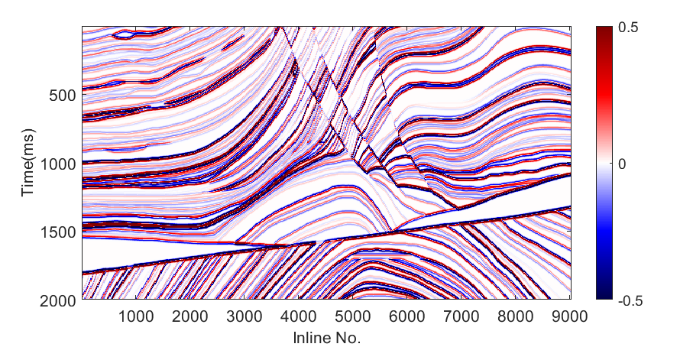}
    \caption{}
  \end{subfigure}
  \begin{subfigure}[b]{0.8\textwidth}
    \includegraphics[width=\textwidth]{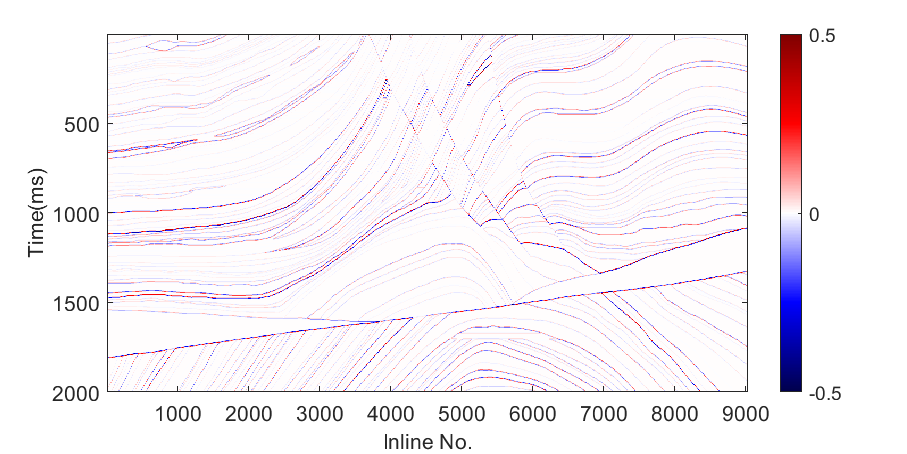}
    \caption{}
  \end{subfigure}
  \begin{subfigure}[b]{0.8\textwidth}
    \includegraphics[width=\textwidth]{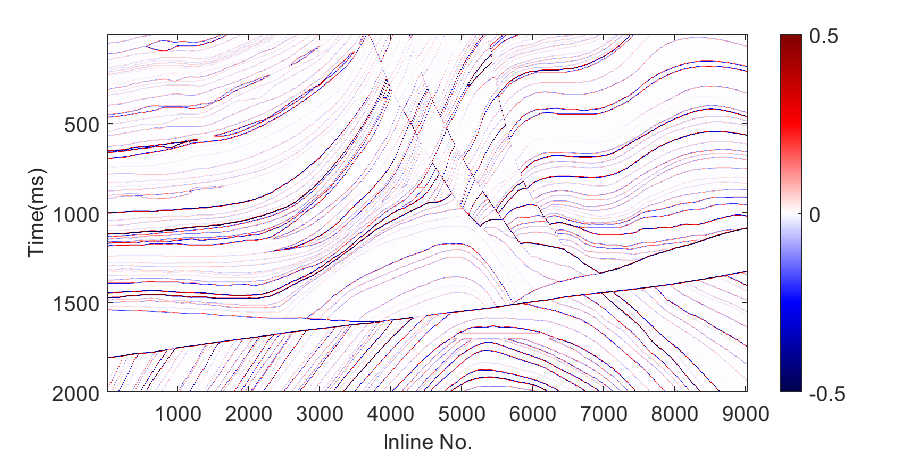}
    \caption{}
  \end{subfigure}
  \caption{Prediction results of DAKD-Net on Marmousi2 model data. (a) Original low-resolution seismic section; (b) Ground-truth high-resolution seismic section; (c) High-resolution prediction result from DAKD-Net.}
  \label{fig:prediction_results}
\end{figure}

To more intuitively demonstrate the spectral extension effect of the student network, a single trace is randomly selected from the prediction results for comparison in both the time and frequency domains. Fig. 10 compares the normalized amplitude spectra of the high-resolution prediction results from DAKD-Net with those of actual high-resolution data and low-resolution data.  It can be observed that the normalized amplitude spectra of the student network's prediction results exhibit higher dominant frequencies and bandwidths compared to the original low-resolution seismic data, and are more closely aligned with the actual high-resolution seismic data. This indicates that the student network can successfully recover the high-frequency components in the seismic data, with the amplitude spectrum improved by approximately two octaves relative to the original data.

\begin{figure}[htbp]
  \centering
  \includegraphics[width=\textwidth]{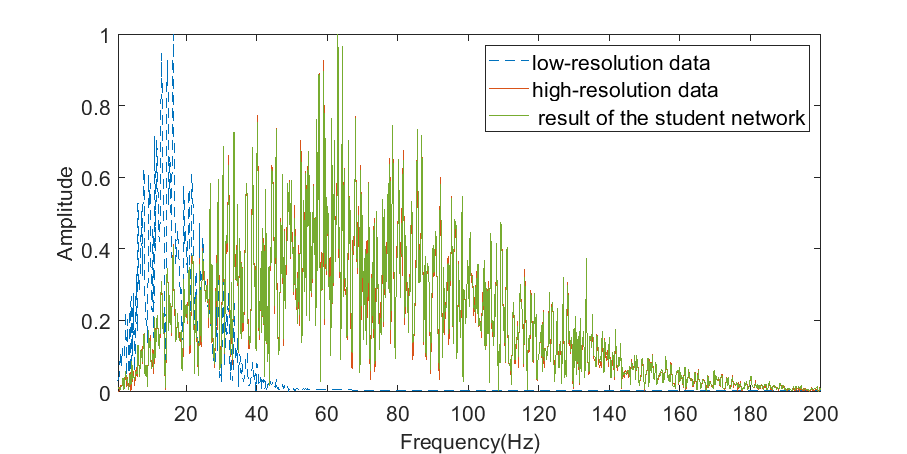}
  \caption{Comparison of normalized amplitude spectra for single-trace data from the Marmousi2 model.}
  \label{fig:amplitude_spectra}
\end{figure}

To demonstrate the advantages of the proposed DAKD-Net method, a FCN was constructed to perform high-resolution prediction on the synthetic data. Its performance was then compared with that of DAKD-Net. FCN is a classical deep learning architecture commonly used for image reconstruction tasks, known for its good feature extraction and reconstruction capabilities. Fig. 11 shows the variation trends of PSNR and SSIM values during the training process for both the traditional FCN method and the DAKD-Net method. As seen in the Fig. 11, although both networks exhibit similar trends in PSNR and SSIM, the PSNR and SSIM values of the FCN are significantly lower than those of DAKD-Net. This indicates that DAKD-Net can more accurately capture detailed and structural information in the data when recovering high-resolution seismic data, whereas the FCN yields comparatively smoother results with lower accuracy in detail recovery.

\begin{figure}[htbp]
  \centering
  \begin{subfigure}[b]{0.45\textwidth}
    \includegraphics[width=\textwidth]{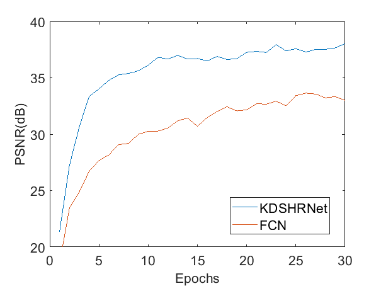}
    \caption{}
  \end{subfigure}
  \begin{subfigure}[b]{0.45\textwidth}
    \includegraphics[width=\textwidth]{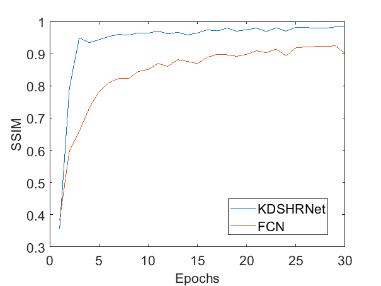}
    \caption{}
  \end{subfigure}
  \caption{Evaluation metrics of FCN and DAKD-Net. (a) PSNR values of FCN and DAKD-Net on the validation set; (b) SSIM values of FCN and DAKD-Net on the validation set.}
  \label{fig:evaluation_metrics}
\end{figure}

Fig. 12 shows the high-resolution prediction results of the two methods, with local areas of the seismic profiles enlarged to highlight their differences. It is evident that in terms of detail recovery, the DAKD-Net method significantly outperforms the traditional FCN method. When predicting seismic profiles, the FCN is able to reconstruct the main structures with relative clarity. However, it shows certain limitations in recovering high-frequency details, with inter-layer information appearing somewhat blurred. In contrast, DAKD-Net delivers superior prediction results. It effectively restores inter-layer information, maintains better continuity of reflection events, and preserves the clarity of high-frequency reflection interfaces. As a result, the overall seismic profile appears richer in detail, with especially notable improvements in areas of complex geological structure.

\begin{figure}[htbp]
  \centering
  \begin{subfigure}[b]{0.45\textwidth}
    \includegraphics[width=\textwidth]{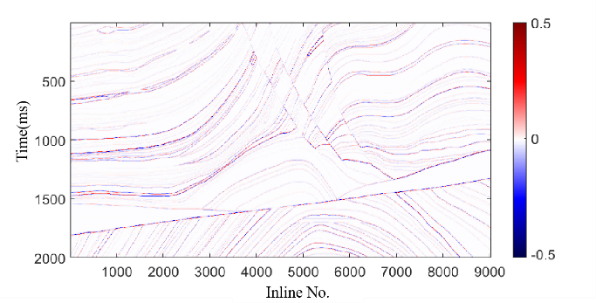}
    \caption{}
  \end{subfigure}
  \begin{subfigure}[b]{0.45\textwidth}
    \includegraphics[width=\textwidth]{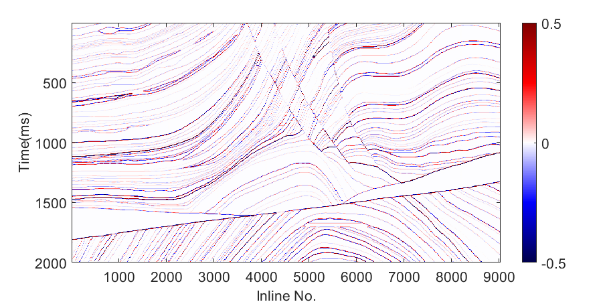}
    \caption{}
  \end{subfigure}
  \begin{subfigure}[b]{0.45\textwidth}
    \includegraphics[width=\textwidth]{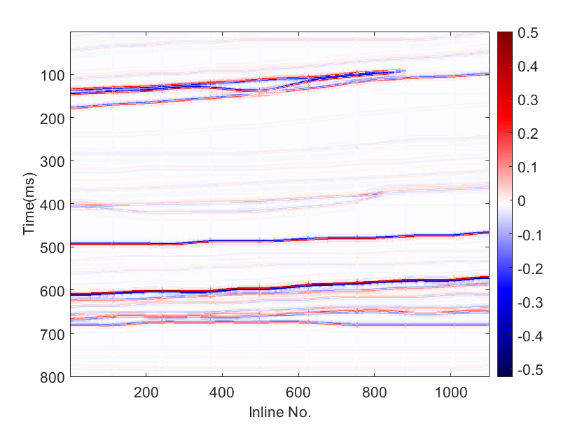}
    \caption{}
  \end{subfigure}
  \begin{subfigure}[b]{0.45\textwidth}
    \includegraphics[width=\textwidth]{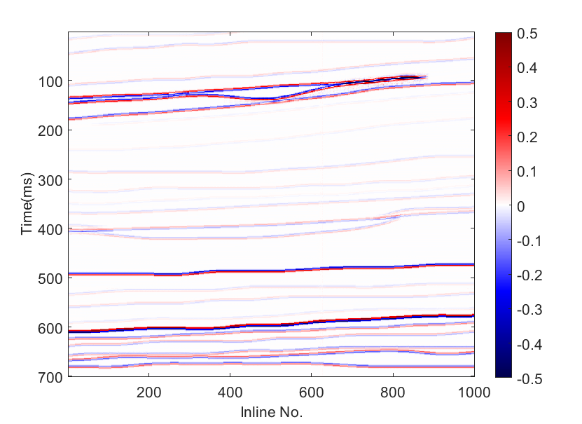}
    \caption{}
  \end{subfigure}
  \caption{Comparison of prediction results between FCN and DAKD-Net. (a) Prediction result of FCN. (b) Prediction result of DAKD-Net. (c) Local region of FCN prediction result. (d) Local region of DAKD-Net prediction result.}
  \label{fig:comparison_results}
\end{figure}

To evaluate the robustness of the DAKD-Net method under different signal-to-noise ratio (SNR) conditions, Gaussian noise with varying variances was randomly added to the Marmousi2 model data to generate noisy datasets with SNR levels of 1 dB, 3 dB, 5 dB, and 10 dB. DAKD-Net was then used to perform high-resolution predictions on the noisy synthetic data. Fig. 13 and Fig. 14 show the noisy data and the corresponding high-resolution prediction results under different SNR conditions. It can be observed that DAKD-Net maintains good robustness under various noise levels, effectively suppressing random noise while achieving high-resolution reconstruction. Overall, the higher the SNR, the better the noise suppression performance. From the comparison, it is evident that as the SNR decreases, the degree of noise interference increases, and the effectiveness of the high-resolution processing is reduced. In particular, when the SNR drops below 3 dB, the influence of noise becomes significantly more pronounced, making it difficult for the high-resolution network to achieve the expected reconstruction accuracy. Under such conditions, it is recommended to apply noise suppression processing before performing high-resolution reconstruction.

\begin{figure}[htbp]
  \centering
  \begin{subfigure}[b]{0.45\textwidth}
    \includegraphics[width=\textwidth]{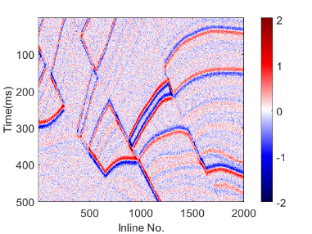}
    \caption{}
  \end{subfigure}
  \begin{subfigure}[b]{0.45\textwidth}
    \includegraphics[width=\textwidth]{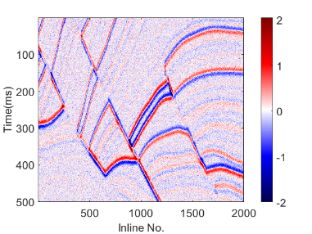}
    \caption{}
  \end{subfigure}
  \begin{subfigure}[b]{0.45\textwidth}
    \includegraphics[width=\textwidth]{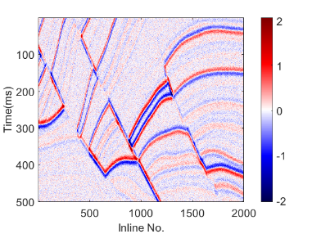}
    \caption{}
  \end{subfigure}
  \begin{subfigure}[b]{0.45\textwidth}
    \includegraphics[width=\textwidth]{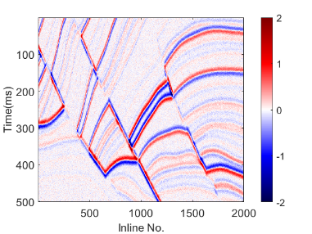}
    \caption{}
  \end{subfigure}
  \caption{Synthetic data with different SNR. (a) SNR = 1 dB; (b) SNR = 3 dB; (c) SNR = 5 dB; (d) SNR = 10 dB.}
  \label{fig:synthetic_snr}
\end{figure}

\begin{figure}[htbp]
  \centering
  \begin{subfigure}[b]{0.45\textwidth}
    \includegraphics[width=\textwidth]{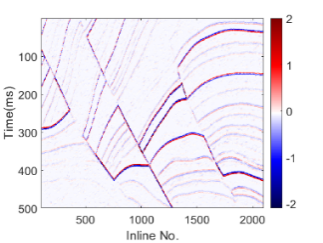}
    \caption{}
  \end{subfigure}
  \begin{subfigure}[b]{0.45\textwidth}
    \includegraphics[width=\textwidth]{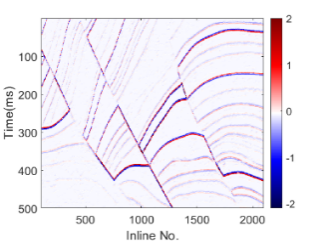}
    \caption{}
  \end{subfigure}
  \begin{subfigure}[b]{0.45\textwidth}
    \includegraphics[width=\textwidth]{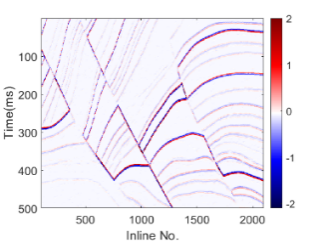}
    \caption{}
  \end{subfigure}
  \begin{subfigure}[b]{0.45\textwidth}
    \includegraphics[width=\textwidth]{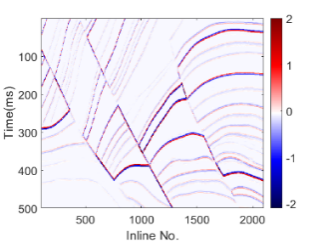}
    \caption{}
  \end{subfigure}
  \caption{High-resolution prediction results under different SNR conditions. (a) SNR = 1 dB; (b) SNR = 3 dB; (c) SNR = 5 dB; (d) SNR = 10 dB.}
  \label{fig:prediction_snr}
\end{figure}

\subsection{Real Seismic Data}
\label{sec8}

The trained model is tested using real seismic data. The actual seismic data is from the Tarim Basin in Xinjiang, specifically from the Tazhong Oil Field. The study area is located in the southwest part of the Akekule uplift, a third-level structural unit on the northeastern sag of the Tazhong Depression. The research target layer is the Middle Triassic oil group, with sand body thickness varying between 0-60m, exhibiting strong geological heterogeneity and significant lateral variation. The actual seismic data has a size of inline 401 * Xline 401 and a sampling rate of 2 ms.

Since the Marmousi2 model is a simulated seismic data model, it can be effectively used to validate and train high-resolution reconstruction methods. However, there are certain differences between its seismic wavelets and frequency characteristics and those of actual field data. These differences cause the prediction results to fall short of expectations when a network trained using the Marmousi2 model is directly tested on real seismic data. To address this issue, a transfer learning approach is introduced. The specific steps are as follows: First, a seismic profile of size Time 500 * Xline 401 in the Inline direction is selected from the actual seismic data. This profile is divided into several 128*128 sized data blocks, from which 20 blocks are randomly selected as low-resolution input data. Then, using the spectral extension method, high-resolution data is constructed from these selected low-resolution data, providing high-precision label data for subsequent high-resolution predictions. A total of 20 pairs of high-resolution and low-resolution seismic data are obtained, which are used as the training dataset for transfer learning fine-tuning. The pre-trained DAKD-Net is fine-tuned using these training datasets by freezing the parameters of the first few layers of the network. Only the last two convolutional layers and residual layers are trained, with their weights updated to enhance the network's generalization ability. A seismic profile that was not involved in the transfer learning process is selected for testing to verify the effectiveness and applicability of this method on actual field data. Fig. 15 presents the high-resolution prediction results of various methods applied to actual seismic data. This includes the original seismic profile (Fig. 15(a)), the predictions from the DAKD-Net method before fine-tuning (Fig. 15(b)), the predictions from the DAKD-Net method after fine-tuning (Fig. 15(c)), and the predictions from the Fully Convolutional Network (FCN) method (Fig. 15(d)).

\begin{figure}[htbp]
  \centering
  \begin{subfigure}[b]{0.45\textwidth}
    \includegraphics[width=\textwidth]{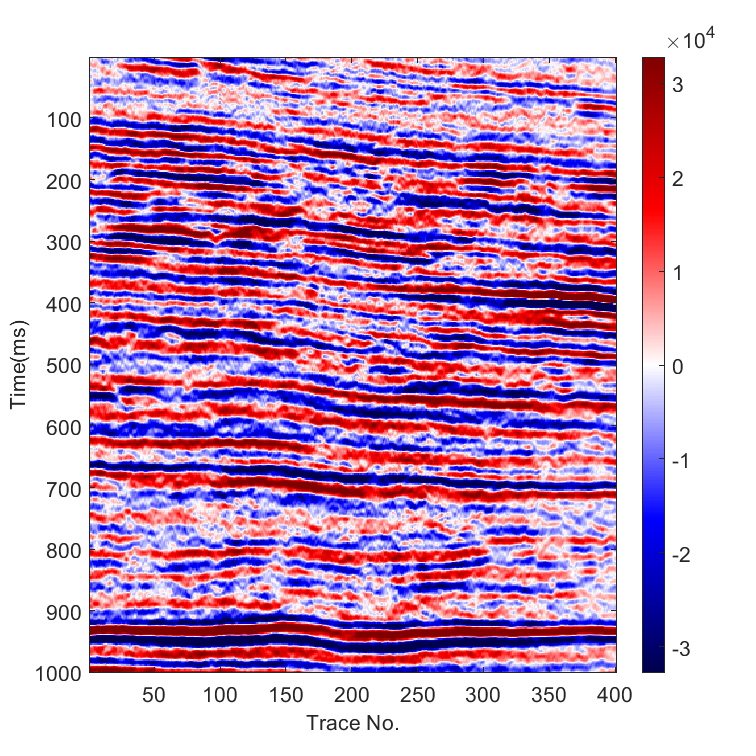}
    \caption{}
  \end{subfigure}
  \begin{subfigure}[b]{0.45\textwidth}
    \includegraphics[width=\textwidth]{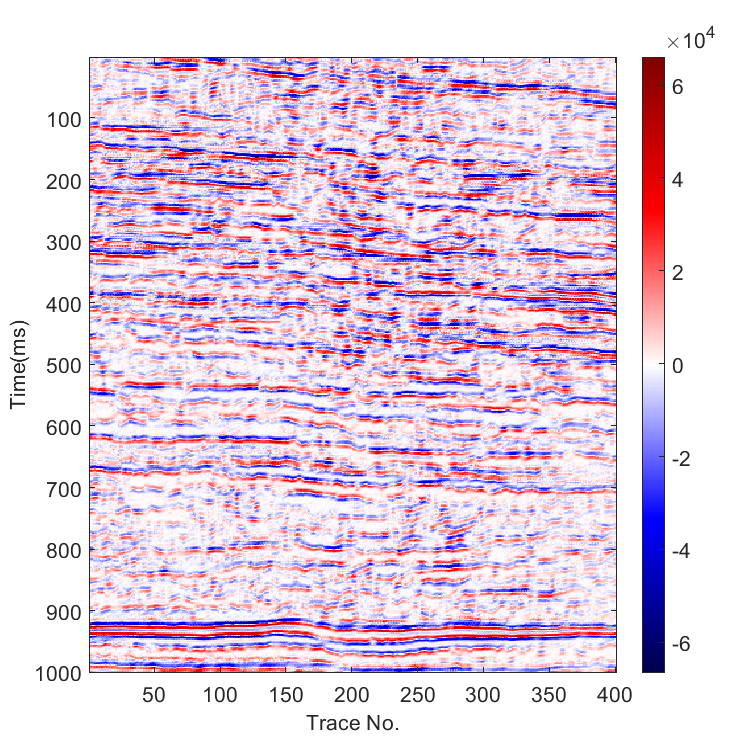}
    \caption{}
  \end{subfigure}
  \begin{subfigure}[b]{0.45\textwidth}
    \includegraphics[width=\textwidth]{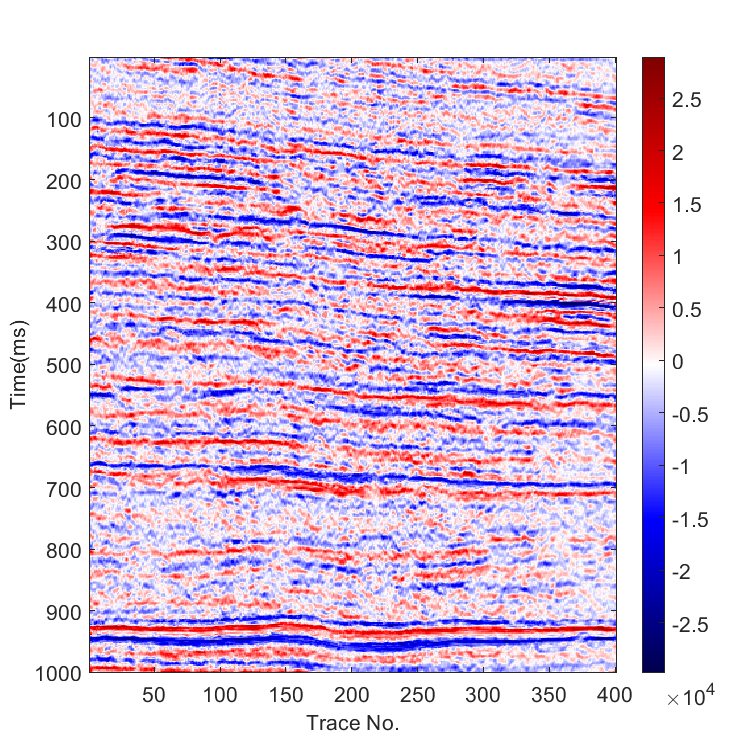}
    \caption{}
  \end{subfigure}
  \begin{subfigure}[b]{0.45\textwidth}
    \includegraphics[width=\textwidth]{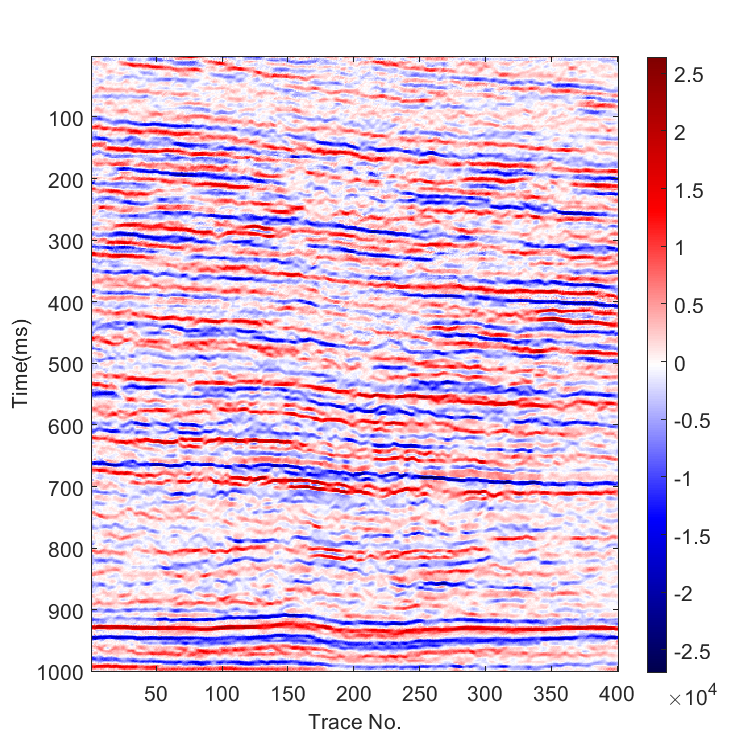}
    \caption{}
  \end{subfigure}
  \caption{Comparison of real seismic profile prediction results. (a) Original seismic profile; (b) DAKD-Net prediction result without transfer learning; (c) DAKD-Net prediction result after transfer learning; (d) FCN prediction result.}
  \label{fig:real_seismic_comparison}
\end{figure}

It can be seen that when the network trained with the Marmousi2 synthetic data is directly applied to high-resolution prediction of actual seismic data, the prediction results show significant deviations. The high-resolution prediction results after fine-tuning via transfer learning are clearly superior to those before fine-tuning. The predicted profile shows more continuous reflection axes, clearer structural features, and better consistency with the actual high-resolution data. When comparing with the FCN method's prediction results, it is observed that the FCN method improves resolution to some extent. However, the continuity of the reflection axes is not as good as that of the high-resolution seismic data network. Some stratigraphic boundaries exhibit blurring and distortion, especially in complex structural areas, where the stability of the prediction results is weaker, as indicated in Fig. 15(c) and 15(d).

A comparison is made between the DAKD-Net method and traditional spectral whitening techniques. Spectral whitening is one of the commonly used methods for high-resolution seismic processing, which aims to flatten the amplitude spectrum within the effective frequency range without altering the phase of the signal, thereby compensating for high-frequency components. The results are illustrated in Fig. 16. The spectral whitening method processes the data on a single trace basis in the frequency domain, neglecting inter-trace relationships. As a result, the entire profile after spectral whitening exhibits noticeable noise, leading to a lack of clarity. In contrast, the DAKD-Net method effectively suppresses noise while preserving more structural details, thereby avoiding the issue of overall enhanced noise that can arise from frequency compensation.

\begin{figure}[htbp]
  \centering
  \begin{subfigure}[b]{0.45\textwidth}
    \includegraphics[width=\textwidth]{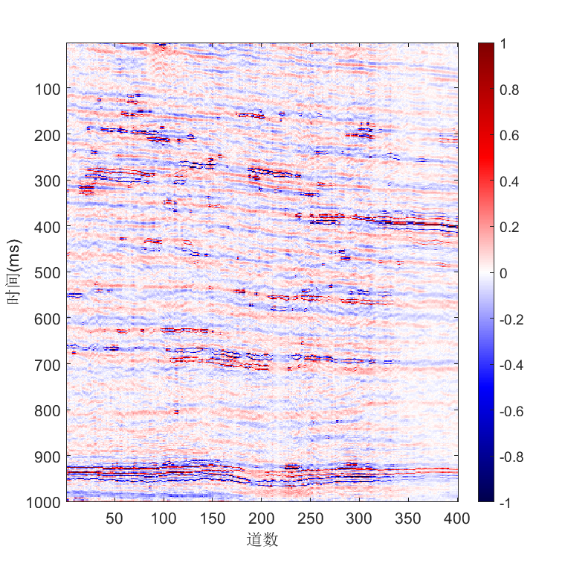}
    \caption{}
  \end{subfigure}
  \begin{subfigure}[b]{0.45\textwidth}
    \includegraphics[width=\textwidth]{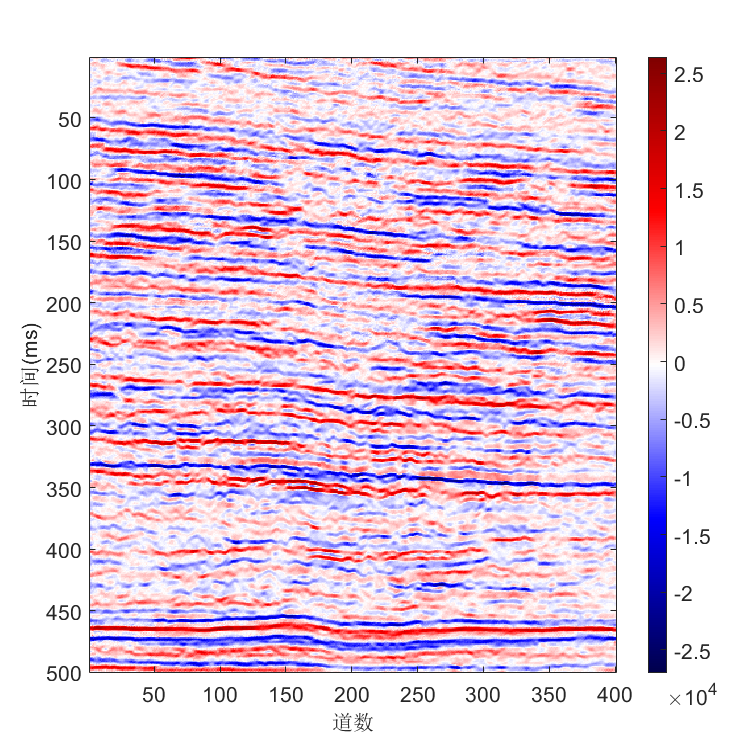}
    \caption{}
  \end{subfigure}
  \caption{Comparison between DAKD-Net and the spectral whitening method. (a) Results from the spectral whitening method; (b) Predictions obtained using DAKD-Net.}
  \label{fig:comparison_spectral_whitening}
\end{figure}

Further comparison of the spectral characteristics between the original seismic data and the high-resolution prediction results is shown in Fig. 17. As illustrated in the figure, the effective frequency band of the original data is mainly concentrated between 20 Hz and 40 Hz, with energy gradually attenuating as frequency increases. After high-resolution processing using DAKD-Net, the data's frequency bandwidth is effectively broadened, and the energy in the high-frequency range is significantly enhanced. The frequency range is expanded to 20 Hz-80 Hz. The dominant frequency is increased, indicating that the network has successfully compensated for the high-frequency components of the seismic data. Moreover, in the low-frequency range, the spectral curve of the predicted data closely matches that of the original data. This demonstrates that while enhancing high-frequency information, DAKD-Net does not introduce additional low-frequency distortion or energy loss.

\begin{figure}[htbp]
  \centering
  \includegraphics[width=\textwidth]{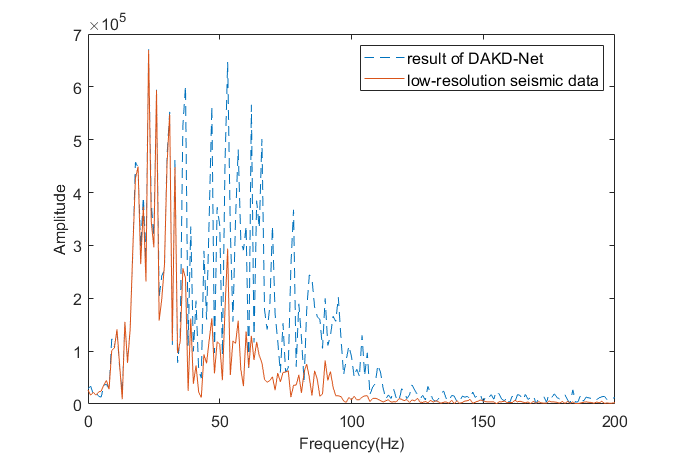}
  \caption{Spectral comparison between the original seismic data and the high-resolution predictions obtained from DAKD-Net.}
  \label{fig:spectral_comparison}
\end{figure}

Finally, the DAKD-Net method was employed for high-resolution processing of three-dimensional seismic data, validated against well log data to analyze the thin layer characterization capability of the seismic data and the well-seismic consistency. Fig. 18 illustrates the frequency enhancement effect on vertical seismic profiles. The red solid line represents the reflection coefficients calculated from well data, convolved with synthetic seismic records generated from different dominant frequency Ricker wavelets. Fig. 18(a) shows the vertical seismic profile extracted from the original three-dimensional data volume, compared with the synthetic seismic record generated using a dominant frequency of 25 Hz Ricker wavelet. Fig. 18(b) displays the seismic profile extracted from the high-resolution processed three-dimensional data volume, overlaid with the synthetic seismic record generated using a dominant frequency of 65 Hz Ricker wavelet. It is evident that the synthetic records correspond well with the seismic profiles, indicating that the high-resolution seismic data post-frequency enhancement can accurately reflect subsurface geological information. The original seismic profile exhibits lower resolution, limiting its capability to characterize thin layers; for instance, two thin layers at a depth of 3040 ms are difficult to identify in the original profile, whereas these layers are clearly visible in the high-resolution profile, demonstrating the method's effectiveness in enhancing thin layer recognition. The high-resolution processed profiles not only exhibit higher vertical resolution, making details clearer, but also show significant improvement in the continuity of the same-phase axes laterally, resulting in clearer inter-layer boundaries and a more coherent overall structure. The reflection waveforms of the synthetic seismic records closely match the processed seismic profiles. These analyses confirm that the DAKD-Net method not only enhances the resolution of seismic data but also preserves the authenticity of the seismic signals, demonstrating a high degree of correlation with well log data. This further validates the effectiveness and reliability of the proposed method.

\begin{figure}[htbp]
  \centering
  \begin{subfigure}[b]{0.8\textwidth}
    \includegraphics[width=\textwidth]{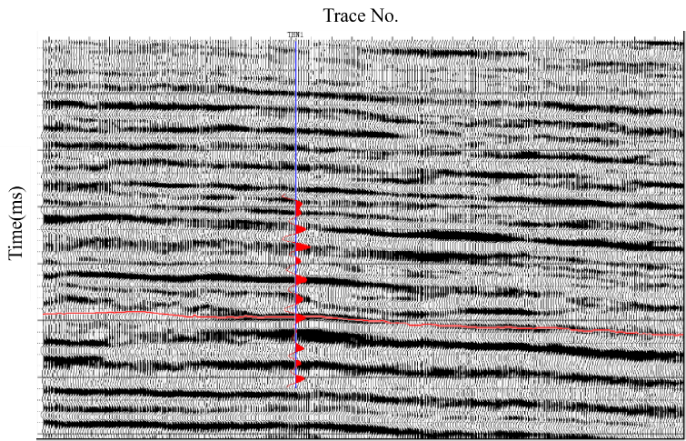}
    \caption{}
  \end{subfigure}
  \begin{subfigure}[b]{0.8\textwidth}
    \includegraphics[width=\textwidth]{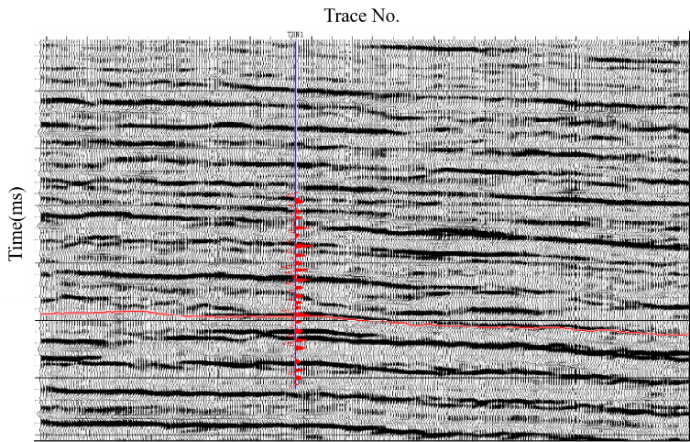}
    \caption{}
  \end{subfigure}
  \caption{Composite record and seismic profile overlay map. (a) Overlay display of original seismic profiles and 25Hz Rayleigh wave composite records; (b) High-resolution processed seismic profiles and 65Hz Rayleigh wavelet composite records are overlaid and displayed.}
  \label{fig:composite_record}
\end{figure}

Fig. 19 compares the changes in root mean square (RMS) amplitude horizontal slices of the 3D seismic data before and after spectral extension. It can be observed that the RMS amplitude of the original data before spectral extension exhibits relatively blurred characteristics, with local details lost. In particular, in areas with complex structures, the amplitude variations are smoother, making boundary identification more difficult. After spectral extension, the RMS amplitude distribution becomes clearer. Local energy is enhanced, making thin-layer features and structural boundaries more prominent. This allows for a more accurate depiction of detailed variations in the seismic data, as well as reservoir boundaries and structural features. These results demonstrate that the spectral extension method significantly improves seismic data resolution and enhances the ability to interpret subsurface geological features.

\begin{figure}[htbp]
  \centering
  \begin{subfigure}[b]{0.45\textwidth}
    \includegraphics[width=\textwidth]{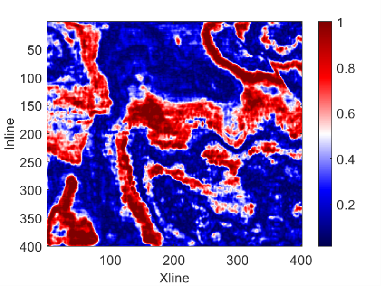}
    \caption{}
  \end{subfigure}
  \begin{subfigure}[b]{0.45\textwidth}
    \includegraphics[width=\textwidth]{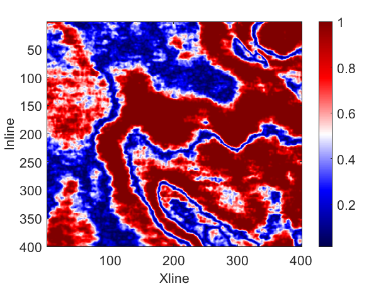}
    \caption{}
  \end{subfigure}
  \caption{RMS amplitude horizontal slice comparison. (a) RMS amplitude horizontal slice before high-resolution processing; (b) RMS amplitude horizontal slice after high-resolution processing.}
  \label{fig:rms_amplitude_comparison}
\end{figure}

\section{Conclusion}
\label{sec:conclusion}

This article presents a deep learning method for enhancing the high-resolution of seismic data, which integrates knowledge distillation and domain adaptation mechanisms—DAKD-Net. The DAKD-Net approach is not limited to single-channel seismic data; it is capable of extracting structural information from inter-channel seismic data, demonstrating higher robustness and efficiency compared to traditional spectral extrapolation methods. By incorporating physical priors through knowledge distillation, DAKD-Net enhances the structural interpretation capabilities relative to conventional Fully Convolutional Networks (FCNs). Additionally, the implementation of a domain adaptation strategy utilizes real high-resolution spectral extrapolation results as supervision, effectively merging physical models with deep features. Experimental results on synthetic datasets and actual field data indicate that the DAKD-Net method significantly improves the clarity of seismic profile details and the continuity of the isochrones, which holds substantial significance for fine reservoir characterization and structural interpretation. Future work will consider the introduction of physics-constrained inversion mechanisms and a broader range of pre-trained models to further enhance the robustness and adaptability of the network.

\section*{Acknowledgements}
This work was supported by the National Natural Science Foundation of China(NSFC) under Grant No.42474168 \& No. 42130812, and. R \& D Center for Ultra-Deep Complex Reservoir Exploration and Development, CNPC and Engineering Research Center for Ultra-deep Complex Reservoir Exploration and Development, Xinjiang Uygur Autonomous Region (Grant No. RDCUD2024-CX01-03) . We are also grateful to the related company for generously providing the field data.

\bibliographystyle{elsarticle-harv} 
\bibliography{referencedp}
\end{document}